\documentclass[10pt,a4paper,final]{article}
\pdfoutput=1
\usepackage[utf8]{inputenc}
\usepackage[T1]{fontenc}
\usepackage[english]{babel}
\usepackage{amsmath}
\usepackage{amsfonts}
\usepackage{amssymb}
\usepackage{graphicx}
\usepackage{kpfonts}
\usepackage[section]{placeins}

\usepackage[left=2cm,right=2cm,top=2cm,bottom=2cm]{geometry}

\usepackage{authblk}
\usepackage{float}

\begin{document}

\title{The role of blood circulatory system in thermal regulation of animals explained by entropy production analysis}

\author{Michael Zakharov \texttt{trapman.hunt@gmail.com}}
\affil{PetPace LTD, Kibbutz Shefayim, Israel}
\author{Michael Sadovsky \texttt{sadovsky.mikhail@gmail.com}}
\affil{Institute of Computational Modelling SB RAS, Russia}

\date{\today}
\maketitle

\begin{abstract}
A novel model of thermal regulation of homoeothermic animals has been implemented. The model is based on a non-equilibrium thermodynamic approach which introduces entropy balance and the rate of entropy generation as a formulation of The Second Law. The model explains linear correlation between an animal's skin and environment temperatures using the first principles and demonstrates the role of blood circulation in the thermoregulation of homoeothermic animals. 
\end{abstract}

\section*{Introduction}
Physically, any living being is a heat engine consuming fuel (food) and an oxidizer (oxygen, commonly) and producing some mechanical work accompanied by production of heat. In such capacity, we exclude plants, mushrooms, etc. the so called autotrophic organisms from consideration.

Keeping the analysis exclusively to the animals, we also should divide them into two basic classes: haematocryal (or poikilothermal) creatures, and homoiothermal creatures. Obviously, the animals from both classes have to produce some heat, as a by-product of the mechanical work being done. The point is that poikilothermal animals (those from the first class) are considered to be in a thermal equilibrium with the environment, while the homoiothermal ones exhibit good maintenance of the body temperature. The difference is crucial, from the point of view of excessive, or ``junk'' heat production.

Any creature producing mechanical work using a fuel oxygenation may not be in absolute temperature equilibrium with the environment. However, the ratio of $H_{\textrm{p}}$ to $H_{\textrm{h}}$ has a figure proximal to $10^{-5} \div 10^{-3}$, where $H_{\textrm{p}}$ ($H_{\textrm{h}}$, respectively) is the excess heat that can be produced by a poikilothermal animal (by a homoiothermal animal, respectively).

The diversity of living conditions observed worldwide is beyond imagination. It seems to be surprising, but it isn't so: the most general biological laws force a need to inhabit every possible space. In spite of the great biological significance of these laws, we will not discuss it here. 

Such dense inhabitation poses another problem in the modelling of thermal regulation of both poikilothermal and homoiothermal animals: the problem of overheating. Indeed, any living being lives within a clearly determined range of environmental conditions, where temperature is one of the most important parameters. Thus, a model of thermal regulation must take into account this problem from both ``ends'' of the viable temperature range.

From the thermodynamics point of view metabolism is a chain of non-equilibrium chemical reactions in which energy is produced and consumed by the body. These reactions generally form a \emph{steady state process} (the process which is not in thermodynamic equilibrium, but has no dependency on time). Steady state is more general than \emph{dynamic equilibrium} because it accepts that some of the processes may be irreversible. If a system is in a steady state then some flows and entropy production may be non zero, while certain properties of the system are unchanging in time \cite{Lebon2007}.

This article describes neither metabolism in general, nor particular mechanisms of energy generation by certain chemical reactions or mechanical work of the muscles. These details can be found elsewhere (good starting points would be \cite{Kondepudi2008}, and \cite{Sousa2008}).

It is a well known fact that a significant part of the generated energy is released in the form of \emph{heat} and part of the heat is not used for any purpose inside the body.

This excess heat must be transferred out of any endothermic animal's body core. The transfer itself must involve minimal extra activity for maximum efficiency. Excess heat should be dissipated as passively as possible. The best candidate would be convection and diffusion into the ambience through the shell (skin) of the animal.

One serious flaw of the present models is that they use Newton's cooling approach to estimate energy balance and/or the temperature of an animal. This would seem to be incorrect because the animal is not cooling, but is actively regulating to keep constant temperature. Newton's cooling may be a proper description of some processes that establish thermal equilibrium, while living systems struggle against thermal equilibrium because it is an euphemism of ``death'' from the point of view of life.

\section{The Model}
The animal model follows the well known ''core and shell'' description of an animal's body \cite{Tracy1972, SCHOLANDER1950a, Bakken1976}. The model describes heat transfer through the shell (skin) of an animal. The importance of this heat dissipation mechanism is based on the heat dissipation limit theory introduced in \cite{Speakman2010}. Other modes of heat dissipation (e.\,g. dog's panting) are not in the scope of model. 

The model describes heat dissipation from an animal that performs at near to rest metabolism, and is situated in an environment with a temperature that is within the thermo neutral zone (TNZ) of the animal. 

\begin{figure}[H]
\centering
\includegraphics[bb=0 0 962 732, viewport=0 0 962 732, clip=true, scale=0.405]{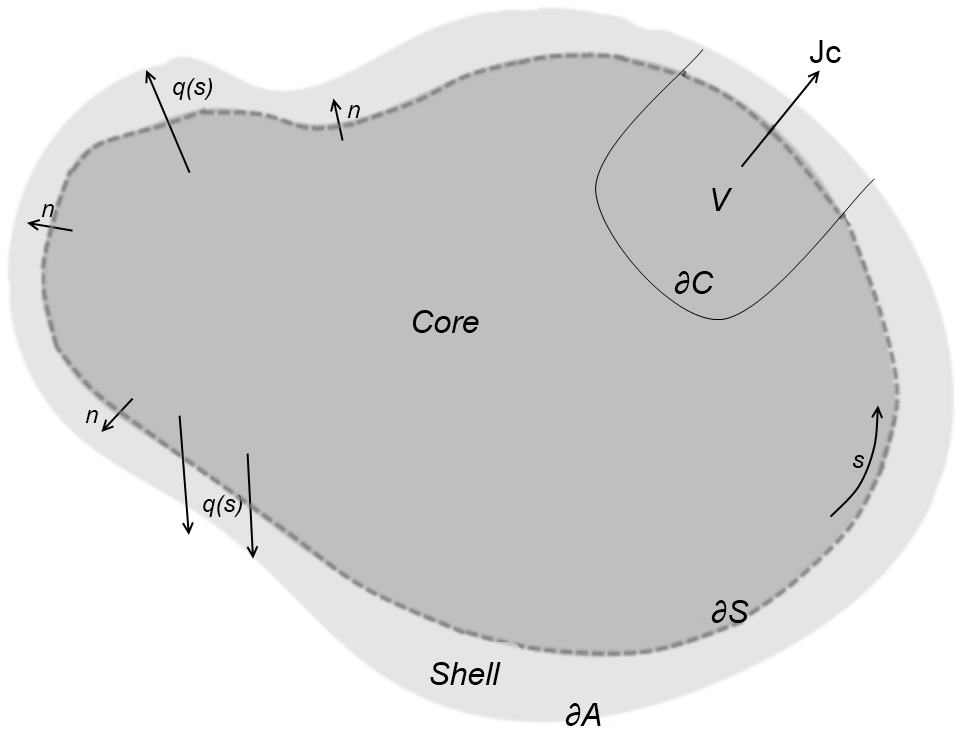}
\caption{Core and Shell Model}
\label{fig:model}
\end{figure}

Major theoretical foundations of heat dissipation by animals can be found in \cite{Wissler1961, Baish1990, Hammel1968}. The model is not related to the animal as a whole, but rather describes local behaviour of heat transfer and possible mechanisms of heat transfer regulation. Animal body is schematically presented in figure~\ref{fig:model}.

\subsection{Core}
The \emph{Core} is the animal body's central compartment, where the heat is generated by metabolic processes. From the point of view of the proposed model the metabolism is a chain of irreversible, non-equilibrium chemical reactions producing energy that is utilized in the body. These reactions generally form a \emph{steady state process} (the non-equilibrium thermodynamic process having no dependency on time).

The major factor of the core functionality model is the well known fact that a significant part of the generated energy is released in form of \emph{heat} and part of the heat is not used for any purposes inside the body. The rate of this heat generation is further noted as \emph{rate of the excessive heat generation}, or $H_{\textrm{exc}}$. Density (volumetric) of $H_{\textrm{exc}}$ is mentioned below as $h_{\textrm{exc}}$. It is important that $H_{\textrm{exc}}$ is produced by an \emph{irreversible process}, i.\,e. metabolism. 

\subsubsection{Assumptions}
The model does not concern itself with internal functionality of the core, but is focused on heat flow from the core to the ambience.

The first assumption is that the \textsl{core is a large, homogeneous closed system and does not have effective spatial structure}. This assumption allows us to consider the core as a system that is entirely in a steady state, and allow us to ignore local (i.\,e. internal to the core) flows, mass transfer, and to use semi-classical approaches for core temperature calculations. This ``large system'' assumption follows the idea that the core has permanent volume, and does not produce any work: $dU_c = dQ_c - p dV_c = dQ_c$, where $dU_c$ is the differential of internal energy of the core.

A further assumption is that the \textsl{core should always} (on the model's time scale) \textsl{be in a steady state}. It means that total entropy production of the core $dS_c=dS_c^{\textrm{ext}} + dS_c^{\textrm{int}}=0$, where $dS_c^{\textrm{int}} \geqslant 0$ due to the Second Law of Thermodynamics. Since internal entropy is generated in an irreversible process (metabolism), all excessive heat should be transferred to the core externals (see details in \cite{Kondepudi2008}):
\begin{equation}
\label{eq:entropy_external}
\begin{split}
&dS_c = dS_c^{\textrm{ext}} + dS_c^{\textrm{int}} = 0 \quad \Rightarrow \quad 
dS_c^{\textrm{ext}} = -dS_c^{\textrm{int}} \\
&dS_c^{\textrm{int}} = \frac{dQ_{\textrm{exc}}}{T_c} \quad \Rightarrow \quad 
dS_c^{\textrm{ext}} = -\frac{dQ_{\textrm{exc}}}{T_c}, \quad \text{or} \\
&\frac{dS_c^{\textrm{ext}}}{dt} = -\frac{1}{T_c}\frac{dQ_{\textrm{exc}}}{dt} = -\frac{1}{T_c}H_{\textrm{exc}}\,;
\end{split}
\end{equation}
it means that core internal energy is constant:
\begin{equation}
\label{eq:energy_internal}
dU_c = dQ_c - p dV_c = dQ_c = T_c dS_c = 0\,,
\end{equation}
and the core temperature $T_c = Q_c/C_v$ does not depend on time:
\begin{equation}
\label{eq:stationary_temperature}
\frac{\partial T_c}{\partial t} = \frac{\partial Q_c}{\partial t} = 0\,.
\end{equation}

Note that the full differential $dQ$ is implemented instead of classical ``imperfect differential'' $\delta Q$ because the model describes steady states of non-equilibrium processes, as opposed to classical descriptions of thermodynamic equilibrium, see \cite{Kondepudi2008, Lebon2007} for detailed explanation.

A specific core temperature $T_c$ is one of the properties of a steady state, and may have a different figure in other steady states. At the same time, metabolism depends on some enzymes whose performance depend (among other factors) on temperature, so keeping $T_c$ inside a preferred temperature range, or even constant, is an advantage. These considerations lead us to the following assumptions of the core model:
\begin{list}{--}{\leftmargin=15mm \labelwidth=5mm \topsep=0mm \labelsep=2mm \itemsep=1pt \parsep=0mm \itemindent=-15pt}
\item core heat production $H_{\textrm{exc}}$ is constant ($h_{\textrm{exc}}$ is a stationary function of coordinates);
\item core temperature $T_c$ is a stationary function of coordinates regardless of external parameters.
\end{list}

Basically, these assumptions (or constraints if we speak about thermoregulation) define simplified thermodynamics of an endothermic animal.

\subsubsection{Formalization}
Entropy balance equation \eqref{eq:entropy_external}, constant internal energy equation \eqref{eq:energy_internal}, core heat production rate $h_{\textrm{exc}}$, and stationary distribution of temperature inside the core \eqref{eq:stationary_temperature} can be combined into a \emph{continuity equation} that is simply another form of the First Law of thermodynamics:
\begin{equation}
\label{eq:continuity_integral_s}
\int_{\partial S}\vec{\textbf{q}}(s) \cdot \vec{\textbf{n}}(s) \, \mathrm{d}S = 
\int_{V_{\textrm{core}}} h_{\textrm{exc}}(\vec{\textbf{r}}) \, \mathrm{d}\vec{\textbf{r}}\,,
\end{equation}
where $\partial S$ is the core surface, $\vec{\textbf{q}}(s)$ is the heat flow, $\vec{\textbf{n}}(s)$ is the outward-pointing unit normal to $\partial S$, $s$ is a parameter that defines coordinates on the surface, $V_{\textrm{core}}$ denotes entire space of the core, $h_{\textrm{exc}}$ is the density of heat generated in the core (see above), and $\vec{\textbf{r}}$ is a coordinate inside the core (figure~\ref{fig:model} illustrates this scheme).

Equation \eqref{eq:continuity_integral_s} simply means that \emph{total outgoing heat flow is equal to the total heat generation rate of the core}. One can apply Gauss theorem (divergence theorem) to re-write~\eqref{eq:continuity_integral_s} as
\begin{equation}
\label{eq:continuity_integral_v}
\int_{V_{\textrm{core}}} \nabla \cdot \vec{\textbf{q}}(\vec{\textbf{r}}) \, \mathrm{d}\vec{\textbf{r}} = 
\int_{V_{\textrm{core}}} h_{\textrm{exc}}(\vec{\textbf{r}}) \, \mathrm{d}\vec{\textbf{r}}\,,
\end{equation}
and write a differential form of the continuity equation for stationary distribution of heat in the core:
\begin{equation}
\label{eq:continuity_differential}
\nabla \cdot \vec{\textbf{q}} = h_{\textrm{exc}}(\vec{\textbf{r}})\,.
\end{equation}
Equation \eqref{eq:continuity_differential} is written for unit volume (in volumetric quantities). This equation is valid for any sub-volume of the core. Thus we can continue with the model applied to an arbitrary volume $V$ that is adjacent to the shell and is separated (conventionally) from the core by surface $\partial C$ as illustrated in figure~\ref{fig:model}.

To have a well-defined physical problem in $V$, we must define the boundary conditions for the equation~\eqref{eq:continuity_differential}. Since we assume constant heat generation rate $\left(\dfrac{\partial h_{\textrm{exc}}}{\partial t}=0\right)\,$ in the core, then the equation \eqref{eq:continuity_integral_s} can be written as
\begin{equation}
\label{eq:continuity_integral_s_const}
\int_{\partial S}\vec{\textbf{q}}(s) \cdot \vec{\textbf{n}}(s) \, \mathrm{d}S = const\,.
\end{equation}
Note that $\vec{\textbf{q}}(s)$ may be a function of time on any given part of the surface $\partial S$, while the whole integral \eqref{eq:continuity_integral_s_const} cannot. We assume that any time dependency in $\vec{\textbf{q}}(s,t)$ is a very slow function:
\begin{equation}\label{eq:weak_time_depend}
\vec{\textbf{q}}(s,dt) = \vec{\textbf{q}}(s,0) + \sum_{n=1}^{\infty} \frac{dt^n}{n!} \, \left. \frac{\partial^n}{\partial t^n}\right|_{t=0}\vec{\textbf{q}}(s,t) = \vec{\textbf{q}}(s,0) + o(dt) \sim \vec{\textbf{q}}(s)\,.
\end{equation}

With this assumption we can finally formulate the local heat balance equation for the core:
\begin{equation}\label{eq:core_heat_balance_flux}
\begin{split}
&\nabla \cdot \vec{\textbf{q}} = h_{\textrm{exc}},\\
&\left. \vec{\textbf{q}} \cdot \vec{\textbf{n}} \, \right|_{\partial S} = J_c\,,
\end{split}
\end{equation}
where all quantities are volumetric, $h_{\textrm{exc}} > 0$, and $J_c \geqslant 0$ by definition and by choice of the coordinate system.

Note that we have not discussed any properties of the flux $\vec{\textbf{q}}$ yet, and the problem~\eqref{eq:core_heat_balance_flux} is a very general representation of the heat balance equation.

\subsubsection{Convection-diffusion equation}
The model provides the description of this flux by the means of a convection-diffusion equation (advection-diffusion equation). The different names for the same equation are used depending on the nature of the subject: ``convection'' is correctly used for vector quantities like magnetic field $\vec{\textbf{B}}$, and ``advection'' is correctly used for scalar quantities like heat $Q$. The difference can be very important under certain circumstances. Nevertheless, it is not important in our case and we continue to use ``convection-diffusion equation'' as a more familiar term.

The model stipulates that $\vec{\textbf{q}}$ is a combination of two types of flux: \emph{diffusive flux} due to thermal diffusion, and \emph{advective flux} which is the flux associated with advection of heat by the flow of fluid:
\begin{equation}
\label{eq:total_flux_def}
\vec{\textbf{q}} = \vec{\textbf{q}}_{dif} + \vec{\textbf{q}}_{adv}\,.
\end{equation}
The model assumes that the diffusive flux is caused by the temperature gradient between the core and shell, and that the convection exists due to the blood flow inside the core and from the core to the shell. 

Blood can work extremely well in advection of heat because the viscosity of blood is~4\,--\,5 times higher than the viscosity of water (including salt water), so the velocity boundary layer exceeds the diffusive boundary layer in greater proportion than that of water (i.\,e. ~Prandtl number $Pr_{\textrm{\,blood}} \sim 30$, while $Pr_{\textrm{water}} \sim 6$ at the typical core temperature $T_{c} \sim 38^\circ \textrm{C}$). Prandtl number is defined as:
\begin{equation}\label{eq:prandtl_number_def}
Pr = \frac{c_p \mu }{k}\,,
\end{equation}
where $c_p$ is the specific heat, $\mu$ is the dynamic viscosity, and $k$ is the thermal conductivity of the fluid. Discussion of exact properties of the blood and blood flow falls beyond the scope of this article and can be found in other researches; see, for example, \cite{Cho2011, Ponder1962, Thurston1972, Hussain1999}.

Advective flux is proportional to the velocity:
\begin{equation}\label{eq:advective_flux_def}
\vec{\textbf{q}}_{adv} = \vec{\textbf{v}} \, Q\,,
\end{equation}
where $Q$ is heat, and $\vec{\textbf{v}}$ is the velocity of the fluid bearing the thermal energy (e.\,g. blood \cite{Wissler1961, Wissler1998}).

Diffusive flux can be approximated by Fourier's law:
\begin{equation}
\label{eq:diffusive_flux_def}
\vec{\textbf{q}}_{dif} = - D \nabla Q\,,
\end{equation}
where $D$ is diffusivity and is defined as \emph{thermal diffusivity} $D = \dfrac{k}{\rho c_v}$ for the heat transfer applications \cite{Venkanna}. 

In this definition, $k$ is the \emph{thermal conductivity} (SI unit of $W{/}(m \cdot K)$), $\rho$ is density, and $c_v$ is \emph{specific heat capacity}. Thermal diffusivity has SI unit of $m^2{/}sec$ and is the material property. Under these approximations the total flux from the core to the shell can be presented as:
\begin{equation}\label{eq:total_flux_final_def}
\vec{\textbf{q}} = - D \nabla Q + \vec{\textbf{v}} \, Q\,,
\end{equation}
and the heat balance equation~\eqref{eq:core_heat_balance_flux} for the flux converts to the equation for the heat~$Q$:
\begin{equation}\label{eq:core_heat_balance_scalar}
\begin{split}
&\nabla \cdot \left(- D \nabla Q + \vec{\textbf{v}} \, Q \right) = h_{\textrm{exc}},\\
&\left. \left(- D \nabla Q + \vec{\textbf{v}} \, Q \right) \cdot \vec{\textbf{n}} \, \right|_{\partial S} = J_c,\\
&\left. Q \, \right|_{\partial C} = f(\vec{\textbf{r}}) = const\,,
\end{split}
\end{equation}
where the additional boundary condition on the (virtual) surface $\partial C$ has been added to make the problem (now the second order PDE) well-defined. 

The boundary condition on $\partial C$ is Dirichlet (the first type) boundary condition. This type of boundary conditions is necessary here to make the solution unique (see \cite{Surhone2010}). The surface $\partial C$ can be an arbitrary one with known temperature distribution, or an isothermal surface, according to suitability, or conforming to the core internal structure (e.\,g. tissues morphology).

The boundary condition on $\partial S$ is Robin (the third type) boundary condition. This type of boundary condition for the convection-diffusion equation is a general form of the \emph{impedance boundary condition} (insulating in case of $J_c = 0$). The latter is quite common in thermodynamic problems, especially for the convective heat exchange applications, see \cite{Surhone2010a} for comprehensive introduction. 

It should be noted that the boundary condition on $\partial S$ is natural and has a meaning of Neumann (the second type) boundary condition in this model, because it simply defines a heat flow across the boundary. The flow has a convective part demonstrating that $\partial S$ is not rigid and has some permeability. 

This boundary position can be estimated using P\'{e}clet number $Pe$. This number is defined as
\begin{equation}
\label{eq:peklet_def}
Pe = \frac{\left| \vec{\textbf{v}} \cdot \nabla T \right|}{\left| k \nabla^2 T \right|} = \frac{L v}{D} = Re \cdot Pr
\end{equation}
for heat transfer applications, where $L$ is a characteristic length, $v$ is a characteristic velocity, $D$ is a thermal diffusivity (as defined above), $Re$ is a Reynolds number, $Pr$ is a Prandtl number. The P\'{e}clet number has the meaning of:
\begin{equation}\label{eq:peklet_meaning}
Pe = \dfrac{\textit{rate of advection}}{\textit{rate of diffusion}}\,.
\end{equation} 
According to this definition, we can say that $\partial S$ lies in a region where the P\'{e}clet number $Pe \sim 1$, or Reynolds number of the blood flow becomes equal to the reciprocal of the blood Prandtl number, $Re \sim 1/Pr$. In that region, the diffusive heat flow becomes more important than the advective heat transfer.

Using known blood flow characteristics (numbers are taken from \cite{Cho2011} and it's citations) and values of thermal properties of animal tissues from \cite{Bowman1975}, we can roughly classify the model regions according to the P\'{e}clet number of the blood flow (see table \ref{tab:peklet_numbers}, where the Prandtl number of the blood $Pr = 30$ is used, see \eqref{eq:prandtl_number_def}).
\begin{table}[H]
\centering
\begin{tabular}{ l c l}
 Region & Reynolds number $Re$ (by order of magnitude) & P\'{e}clet number $Pe$\\
 \hline
 Core vascular & 100 - 1000 & 3000 - 30000 \\
 Subcutaneous vascular & 1 - 10 & 30 - 300 \\
 Microvascular & 0.001 & 0.03 \\
 \hline
\end{tabular}
\caption{P\'{e}clet numbers}
\label{tab:peklet_numbers}
\end{table}

\begin{figure}[H]
\centerline{
\includegraphics[bb=0 0 639 191, viewport=0 0 639 191, clip=true, scale=0.645]{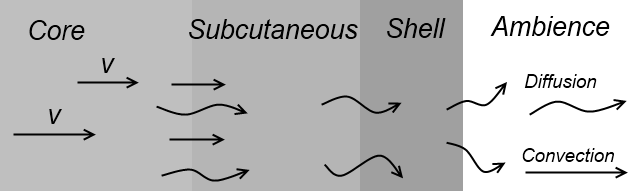}}
\caption{Illustration of $\partial S$ position}
\label{fig:subcutaneous_model}
\end{figure}

The boundary $\partial S$ lays inside the subcutaneous vascular system, close to the internal boundary of the capillary network, as figure \ref{fig:subcutaneous_model} shows. 

In a simple environment where the diffusion coefficient is constant and the flow is incompressible the equation \eqref{eq:core_heat_balance_scalar} can be simplified to:
\begin{equation}\label{eq:core_heat_balance_incompr}
\begin{split}
&- D \nabla^2 Q + \vec{\textbf{v}} \cdot \nabla Q = h_{\textrm{exc}},\\
&\left. \left(- D \nabla Q + \vec{\textbf{v}} \, Q \right) \cdot \vec{\textbf{n}} \, \right|_{\partial S} = J_c,\\
&\left. Q \, \right|_{\partial C} = f(\vec{\textbf{r}}) = const\,,
\end{split}
\end{equation}
or, using core temperature $T_c$ defined from $Q = \rho c_v T_c$ and assuming that $c_v$ and $\rho$ are both constant:
\begin{equation}\label{eq:core_heat_balance_temperature}
\begin{split}
&-k \nabla^2 T_c + \rho \vec{\textbf{v}} \cdot c_v \nabla T_c = h_{\textrm{exc}}, \\
&\left.\left(-k \nabla T_c + \rho \vec{\textbf{v}} c_v T_c \right) \cdot \vec{\textbf{n}} \, \right|_{\partial S}= J_c,\\
&\left. T_c \, \right|_{\partial C} = g(\vec{\textbf{r}}) = const\,,
\end{split}
\end{equation}
where all the quantities are volumetric, $T_c$ is the core temperature, $k $ $[J{/}(sec \cdot m \cdot K)]$ is the thermal conductivity, $\rho $ $[kg{/}m^3]$ is the density, $c_v $ $[J{/}(kg \cdot K)]$ is the specific heat capacity, and $\vec{\textbf{v}}$ $[m{/}sec]$ is the blood flow velocity. All these properties may be determined either by a tissues, or by the blood, because the core tissues are well perfused and it is possible to consider the tissue and vasculature as a whole for heat transfer models \cite{Baish1986, Baish1990}. 

Both $J_c$ and $h_{\textrm{exc}}$ are locally defined and depend on the type and size of the animal (Klieber's law \cite{kleiber6body, Savage2004}), the current activity level of the animal, the animal's fitness level, etc. Problem \eqref{eq:core_heat_balance_temperature} can be interpreted as a formalization of the homeostasis property of endothermic animals.

The heat flow $J_c$ to the ambience, metabolic heat production rate $h_{\textrm{exc}}$, and core temperature $T_c$ define a steady state of the core through equation~\eqref{eq:core_heat_balance_temperature}. In this equation $h_{\textrm{exc}}$ is ``inhomogeneous right side'' of the convection equation, and $J_c$ along with $T_c$ are members of the boundary conditions. 

\subsubsection{Core entropy balance estimation}
If the core (which can be assumed to be a large system in equilibrium for the illustrative purposes) were connected to an ambience (another large system in equilibrium), we could estimate the entropy production in the core and the ambience in the following way: the core and the ambience heat exchange can be presented as a heat conduction in an isolated system consisting of two parts, where each part is in equilibrium and has a well-defined temperature $T_c$ and $T_a$ (assuming $T_c > T_a\,$), and constant volume. The mass exchange can be neglected since the core is said to be a closed system. 

This system will maintain a heat flow $J_Q$ that transfers $dQ$ per time $dt$ per unit area from the hotter part to the colder one. Since the volume is constant, the energy changes only by the heat flow and we can write $dU=dQ$ for each part of the system. The First Law yields $-dQ_c=dQ_a=dQ$, thus the total change of entropy in this system is
\begin{equation}
\label{eq:core_entropysimple}
dS=-\frac{dQ}{T_c}+\frac{dQ}{T_a}=\left(\frac{1}{T_a}-\frac{1}{T_c}\right)dQ\,.
\end{equation}

Since $T_c > T_a$, one can see that $dS>0$. Finally, we can write an expression for the \emph{rate of entropy production} as
\begin{equation}
\label{eq:core_entropyratesimple}
\dot{S}=\left(\dfrac{1}{T_a}-\dfrac{1}{T_c}\right)\dfrac{dQ}{dt}=\left(\dfrac{1}{T_a}-\dfrac{1}{T_c}\right)J_c\,,
\end{equation}
where $J_c\equiv\frac{dQ}{dt}$ is heat flow per unit area from the boundary conditions of the problem \eqref{eq:core_heat_balance_scalar}. Unfortunately, this approach is not appropriate because the core is connected to the shell and the shell may not be in the thermodynamic equilibrium.

\subsection{Shell}
The model describes the shell as a system that is confined between two ``large'' systems: the core and the ambience. Both systems have certain temperatures ($T_c$ and $T_a < T_c$). There is a heat flow $\vec{\textbf{J}}_Q(\vec{\textbf{r}}, t)$ through the shell transferring the heat from the core to an ambience. The shell is in a steady state most of the time, but it can also pass through some transitive states.

\subsubsection{Assumptions}
We use the following assumptions, most of which are common assumptions of non-equilibrium thermodynamics:
\begin{list}{--}{\leftmargin=15mm \labelwidth=5mm \topsep=0mm \labelsep=2mm \itemsep=1pt \parsep=0mm \itemindent=-15pt}
\item all flows are small;
\item variations of flows are small (i.\,e. gradients and higher order derivatives can be neglected);
\item all thermodynamic forces variations are small;
\item all of the above are true for any part of the shell and are true for the entire shell (i.\,e. the shell is homogeneous and has ``small'' thickness).
\end{list}
These assumptions can be combined into just one: the shell conforms to the local equilibrium requirements (see details in \cite{Onsager1931, Kondepudi2008, Lebon2007}). We also assume that the shell is ``passive'' and it's metabolism can be neglected (i.\,e. the shell does not have internal heat sources).

All of the assumptions proposed above seem to be valid and reasonable for the temperature and energy ranges that a living creature may normally encounter. 

The shell is in permanent contact with ambience. The ambient properties may change at any moment in ways that are completely uncontrollable. For this reason we \textsl{do not assume that the shell is in a steady state permanently}. Transient processes are allowed for and will be briefly evaluated.

The shell moves into transitive state immediately after the ambient temperature or any other thermal parameter of the ambience has changed. 

\subsubsection{Formalization}
We start from a derivation of the equation for the temperature distribution in the shell when it is in a steady state. As we already know from the equations \eqref{eq:entropy_external} and \eqref{eq:energy_internal}, steady state means a stationary distribution of temperature as is formalized by equation \eqref{eq:stationary_temperature}.

It yields the following equation that is equivalent to the equation \eqref{eq:continuity_differential} while written for $\vec{\textbf{J}}_Q$:
\begin{equation}\label{eq:shell_continuity_differential}
\nabla \cdot \vec{\textbf{J}}_Q = 0\,,
\end{equation}
where $\vec{\textbf{J}}_Q$ is the heat flow that traverses the shell, and $\vec{\textbf{r}}$ is the coordinate in the shell. The equation is homogeneous because we assume that the shell does not contain any heat sources.

The blood inside of the shell flows in capillaries and it has a velocity that is lower by orders of magnitude than the blood flow velocity in the core, and according to figures from the table \ref{tab:peklet_numbers} the P\'{e}clet number for blood flow in capillaries is $Pe \sim 0.03$.

Thus, we believe that $\vec{\textbf{J}}_Q$ is diffusive and can be approximated in the same way as in equation~\eqref{eq:diffusive_flux_def}:
\begin{equation}\label{eq:shell_diffusive_flux_def}
\vec{\textbf{J}}_Q = - D_s \nabla Q\,,
\end{equation}
where $D_s = \dfrac{k_s}{\rho_s c_s}$ is the thermal diffusivity coefficient for the shell. Since we assume a homogeneity for the shell, then $D_s$ is a function neither of time, nor of coordinates, then \eqref{eq:shell_continuity_differential} can be transformed into:
\begin{equation}\label{eq:shell_heat_balance_scalar}
- D_s \nabla^2 Q =0\,,
\end{equation}
or, taking temperature from $Q = \rho_s c_s T_s$, one can write:
\begin{equation}\label{eq:shell_heat_balance_temp}
\nabla^2 T_s = 0 \,,
\end{equation}
where $T_s$ is the shell temperature, and $k_s$ is the thermal conductivity of the shell.

We need some boundary conditions to complete the physical problem definition. The model proposes Dirichlet (the first type) boundary conditions on the outer surface $\partial A$, and Neumann (the second type) boundary conditions on the inner surface $\partial S$. The final equation together with its boundary conditions looks similar to \eqref{eq:core_heat_balance_temperature}, but does not include convection:
\begin{equation}\label{eq:shell_heat_balance_temperature}
\begin{split}
&\nabla^2 T_s = 0 \,, \\
&\left. k_s \nabla T_s \cdot \vec{\textbf{n}} \right|_{\partial S} = -J_Q\,, \\
&\left. T_s \right|_{\partial A} = T_{\textrm{outer}}\,.
\end{split}
\end{equation}

The boundary conditions on the inner surface $\partial S$ simply mean a heat flow continuity. The boundary conditions on the outer surface $\partial A$ introduce a new temperature $T_{\textrm{outer}}$. We may use a common convective heat transfer approximation (Newton's law of cooling) and stipulate that $T_{\textrm{outer}}$ is defined by the equation $J_c = \varkappa (T_{\textrm{outer}} - T_a)$ (per unit area), where $\varkappa$ is the heat transfer coefficient. Alternatively, we calculate $T_{\textrm{outer}}$ from the entropy production rate of the system and demonstrate that convective approximation is simply a special case of the more general dependency.
 
\subsubsection{Shell entropy production}
The equation for the entropy production rate can be written using the same method as is illustrated during evaluation of \eqref{eq:core_entropyratesimple}, but the equation should be generalized as an equation for production rate of \emph{entropy density}, $\sigma(\vec{\textbf{r}}, t)$ in every elementary volume $\mathrm{d}\vec{\textbf{r}}$, because we can assume only the local equilibrium. Taking note that $\left(\frac{1}{T_a}-\frac{1}{T_c}\right)$ becomes $\nabla T^{-1} $, one can write:
\begin{equation}\label{eq:shell_entropyrate_density}
\sigma(\vec{\textbf{r}}, t)\,\mathrm{d}\vec{\textbf{r}} = \nabla \frac{1}{T(\vec{\textbf{r}})} \cdot \vec{\textbf{J}}_Q(\vec{\textbf{r}}, t)\,\mathrm{d}\vec{\textbf{r}} \, ,
\end{equation}
where $\sigma(\vec{\textbf{r}}, t)$ is the production rate of entropy density, $T(\vec{\textbf{r}})$ is the temperature of the shell, $\vec{\textbf{J}}_Q(\vec{\textbf{r}}, t)$ is a heat flow that ``traverses'' the shell, $\vec{\textbf{r}}$ is the coordinate in the shell.

One can integrate \eqref{eq:shell_entropyrate_density} over the shell and obtain the total entropy production rate in the shell as
\begin{equation}\label{eq:shell_entropyrate_full_3D}
\dot{S}_{\textrm{shell}}(t) = \int_V \sigma(\vec{\textbf{r}}, t)\,\mathrm{d}\vec{\textbf{r}} = \int_V \vec{\textbf{J}}_Q(\vec{\textbf{r}}, t) \cdot \nabla \frac{1}{T(\vec{\textbf{r}})} \,\mathrm{d}\vec{\textbf{r}} \, ,
\end{equation}
where $\int_V \,\mathrm{d}\vec{\textbf{r}}\,$ means an integration over the entire shell volume.

\begin{figure}[H]
\centering
\includegraphics[bb=0 0 765 440, viewport=0 0 765 440, clip=true, scale=0.55]{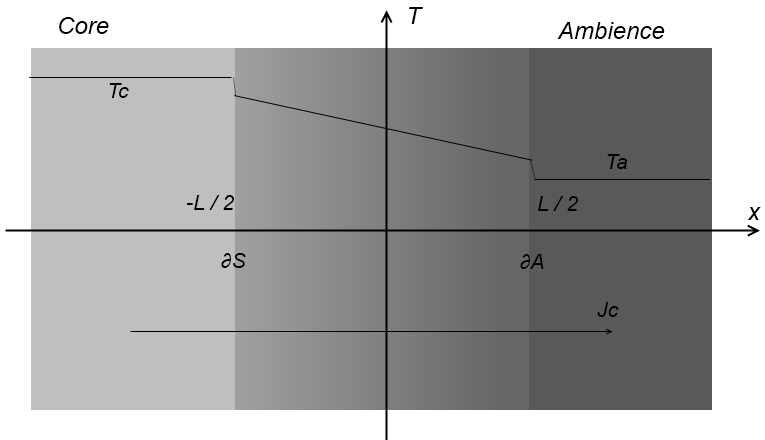}
\caption{Shell Temperature Model}
\label{fig:shell_temperature_model}
\end{figure}

For the sake of clear demonstration of the major properties of the system, we consider the unit area cross section of the shell and assume that the shell boundaries are perpendicular to the heat flux $\vec{\textbf{J}}_Q$. An illustration of this model is presented in figure \ref{fig:shell_temperature_model}.

This simplification removes the routine vector calculus transformations of the volume integral in~\eqref{eq:shell_entropyrate_full_3D} and reduces the 3D~equation space to one dimension.

We can re-write \eqref{eq:shell_entropyrate_full_3D} as a one-dimensional linear integral. The shell cross section area is omitted because we use per unit area values (and volumetric units as usual):
\begin{equation}\label{eq:shell_entropyrate_full_1D}
\dot{S}_{shell}(t) = \int\limits_{-L/2}^{L/2} J_Q(x,t) \frac{d(1/T(x))}{dx}\,\mathrm{d}x\, ,
\end{equation}
where $L$ is the shell thickness, and $x$ is the coordinate across the shell. The origin of the coordinate system is set at the shell median plane and the positive direction points to the ambience. In these coordinates the core surface $\partial S$ is at $x=-L/2$ and the ambient surface $\partial A$ is at $x=L/2$, see figure~\ref{fig:shell_temperature_model} for the illustration.

Integral \eqref{eq:shell_entropyrate_full_1D} can be integrated by parts and evaluates to:
\begin{equation}\label{eq:shell_entropyrate_I1}
\begin{split}
&\dot{S}_{\textrm{shell}}(t) = \left.\dfrac{1}{T(x)} J_Q(x,t)\right|_{-L/2}^{L/2} - \int\limits_{-L/2}^{L/2} \dfrac{1}{T(x)} \dfrac{d J_Q(x,t) }{dx}\,\mathrm{d}x = \dot{S}_{\textrm{steady}}(t) + \dot{S}_{\textrm{transient}}(t)\,, \\
&\dot{S}_{\textrm{steady}}(t) = \dfrac{J_a(t)}{T_a} - \dfrac{J_c(t)}{T_c}\,,\\
&\dot{S}_{\textrm{transient}}(t) = -\int\limits_{-L/2}^{L/2} \dfrac{1}{T(x)} \frac{d J_Q(x,t) }{dx}\,\mathrm{d}x\, ,
\end{split}
\end{equation}
where $J_a(t)=J_Q(L/2,t)$ is the heat flux to the ambience, $J_c(t)=J_Q(-L/2,t)$ is the heat flux from the core, $\dot{S}_{\textrm{steady}}(t)$ is the steady state contribution, and $\dot{S}_{transient}(t)$ is the \emph{transient process contribution}. Equations~\eqref{eq:shell_entropyrate_I1} immediately lead to the steady state entropy production rate formula:
\begin{equation}\label{eq:shell_entropyrate_steady}
\dot{S}_{\textrm{steady}} = \frac{J_a(t)}{T_a} - \frac{J_c(t)}{T_c} = \left(\frac{1}{T_a}-\frac{1}{T_c}\right)J_c\,, 
\end{equation}
under the local equilibrium assumption $\dfrac{d}{dx} J_Q(x,t) = 0$ in any steady state, and we have $J_a(t) = J_c(t) = J_c$ (strictly speaking, on average), and $J_c$ is not a function of time. It is clear that $\dot{S}_{\textrm{steady}} > 0$ if $T_c > T_a$. The equation~\eqref{eq:shell_entropyrate_steady} demonstrates a very important model limitation: inequality $T_c \geqslant T_a$ must be valid at any time. 

The transient process contribution $\dot{S}_{\textrm{transient}}$ can be estimated using local equilibrium assumptions. First, expand integrand into Taylor series around $x=0$:
\begin{equation}\label{eq:shell_entropyrate_I2}
\dot{S}_{\textrm{transient}}(t) = -\int\limits_{-L/2}^{L/2} 
\sum_{n=0}^{\infty} \dfrac{x^n}{n!} \dfrac{\mathrm{d}^n}{\mathrm{d}x^n} \left.\left(\dfrac{1}{T(x)} \dfrac{d J_Q(x,t) }{dx}\right) \right|_{x=0} \,\mathrm{d}x\,,
\end{equation}
where $\dfrac{\mathrm{d}^0}{\mathrm{d}x^0}f(x)\equiv f(x)$, change the order of summation and integration, take note that all odd numbered members of the series are zero because of the integration limits symmetry, neglect any high (i.\,e. $\geqslant 2$) order derivatives by the accepted assumptions, and transform equation~\eqref{eq:shell_entropyrate_I2} into:
\begin{equation}\label{eq:shell_entropyrate_I4}
\dot{S}_{\textrm{transient}}(t) = - L \left.\left(\frac{1}{T(x)} \frac{d J_Q(x,t) }{dx}\right) \right|_{x=0} =
- L \frac{1}{T_s(t)} \frac{d }{dx} J_Q(t)\,,
\end{equation}
where the shell temperature is introduced as $T_s(t) \equiv T(0,t)$. Finally, the equation~\eqref{eq:shell_entropyrate_I4} can be expressed as
\begin{equation}\label{eq:shell_entropyrate_transient}
\dot{S}_{\textrm{transient}}(t) = -\frac{J_a(t) - J_c(t)}{T_s(t)}\,,
\end{equation}
where $\dfrac{d }{dx} J_Q(t)$ has been estimated as $\dfrac{J_a - J_c}{L}$\,, ignoring high order derivatives. 

It is important that one should not use stationary flows in \eqref{eq:shell_entropyrate_transient}, because this equation describes non-steady states, where both $J_a$ and $J_c$ may be functions of time. The same is true for the shell temperature $T_s$. Complete estimation of total entropy production rate in the shell is then: 
\begin{equation}\label{eq:shell_entropyrate_full}
\dot{S}_{\textrm{shell}}(t) = \frac{J_a(t)}{T_a} - \frac{J_c(t)}{T_c} - \frac{J_a(t) - J_c(t)}{T_s(t)}\,,
\end{equation}
where both $T_a$ and $T_c$ are assumed to be weakly dependent on time (at most) in the same sense as explained by~\eqref{eq:weak_time_depend}.

\subsubsection{Shell temperature estimation}
Since the shell temperature $T_s$ appears only in equation~\eqref{eq:shell_entropyrate_transient} for a transient process, we will derive it's value from the transient process properties. The shell can be set into a transient state by the ambient temperature variation. 

If the ambient temperature $T_a$ changes then the shell would move to another steady state according to the steady state entropy production equation~\eqref{eq:shell_entropyrate_steady}. The new state depends on a new ambient temperature, $\tilde{T}_a = T_a + \delta T_a$. Because temperatures of both core and ambience, and heat flow $J_c$ are all independent parameters of the steady state, we can choose $J_c = const$. The choice is quite logical since $J_c$ takes its origin in the core, and nothing can influence the latter to change $J_c$ at the beginning of the transient process in the shell.

The transition process takes some time $\tau$. During this time the shell entropy production rate is different and the excessive heat generation rate is equal to a value set by equation~\eqref{eq:shell_entropyrate_transient}. 

Explanations of the reasons why the shell would move to a transient state and why it will find another steady state, as well as a good estimation of $\tau$ are far beyond the scope of this article.

A total excessive entropy generated by the transient process during its life time $\tau$ is equal to the integral of the entropy production rate over time:
\begin{equation}\label{eq:shell_entropy_transient}
\delta S_{\textrm{transient}} = - \int\limits_0^\tau \frac{J_a(t) - J_c}{T_s(t)}\,\mathrm{d}t\,.
\end{equation}
Using the linear approximations for $T_s(t)$ and $J_a(t)\,$, one can write:
\begin{equation}\label{eq:shell_flux_linear_approx}
\begin{split}
T_s(t) &= T_s(0) + t \left.\frac{\mathrm{d}T_s}{\mathrm{d}t}\right|_{0}\\
J_a(t) &= J_a(0) + t \left.\frac{\mathrm{d}J_a}{\mathrm{d}t}\right|_{0}\,,
\end{split}
\end{equation}
and equation \eqref{eq:shell_entropy_transient} transforms to:
\begin{equation}\label{eq:shell_entropy_transient_linear_approx}
\delta S_{\textrm{transient}} = - \int\limits_0^\tau \dfrac{J_a(0) + t\left.\dfrac{\mathrm{d}J_a}{\mathrm{d}t}\right|_{0} - J_c}{T_s(0) + t\left.\dfrac{\mathrm{d}T_s}{\mathrm{d}t}\right|_{0}}\,\mathrm{d}t\,,
\end{equation}
and after some transformations borrowed from the calculus hand-book, and application of the assumption \\ $T^{-1}_s(0)\left.\dfrac{\mathrm{d}T_s}{\mathrm{d}t}\right|_{0}\tau \ll 1$ has become:
\begin{equation}\label{eq:shell_excess_entropy}
\delta S_{\textrm{transient}} \approx \dfrac{J_c - J_a(0)}{T_s(0)} \tau = - \dfrac{\delta J_a}{T_s(0)} \tau\,,
\end{equation}
because $J_a(0)$ is the heat flow to the ambience at the beginning of the transient process and can be expressed as $ J_a(0) = J_c + \delta J_a$ (in steady state $J_a = J_c$, or $J_a(-0) = J_c $).

An excess entropy given by equation~\eqref{eq:shell_excess_entropy} makes the shell gain (or lose) an amount of heat approximately equal to
\begin{equation}\label{eq:shell_excess_heat}
\delta Q_{\textrm{shell}} = - \tau \delta J_a\,.
\end{equation}
Thus, variation of the shell temperature is equal to
\begin{equation}\label{eq:shell_temperature_varJ}
\delta T_s = - \dfrac{\tau}{\rho_s c_s} \delta J_a\,,
\end{equation}
where the equation is conveniently written in per unit volume, per unit square units, and $\rho_s$ is the shell density, $c_s$ is specific heat capacity of the shell.

Equation~\eqref{eq:shell_temperature_varJ} connects the shell temperature variation to the variation in heat flow to the ambience. A notable property of this equation is the opposite direction of these variances: if the flow increases at the beginning of a transient process ($ \delta J_a > 0 $) then the shell temperature decreases by the end of the transition ($ \delta T_s < 0$), and \textit{vice versa}. This is true for any type of flow and is not related to the flow dependencies on $T_a$ and $T_s$ (as long as the model assumptions are valid). 

We can estimate the transient process time $\tau$ in a simple situation, where the heat flow to the ambience $J_a$ can be presented as a linear combination of conduction, convection, and radiation flows, and heat transfer from the shell to the ambience can be described using the Newton's law of cooling. Newton's law of cooling application assumes that the object (the shell in our case) has a high heat capacity and a high temperature conductivity. Such an assumption is more or less valid for sufficiently large animals.
\begin{equation}\label{eq:shell_simple_Ja}
J_a = J_{\textrm{conduction}} + J_{\textrm{convection}} + J_{\textrm{radiation}}\,.
\end{equation}

It is possible to use linearisation by temperature and express all of them collectively as
\begin{equation}\label{eq:shell_simple_Ja_linear}
J_a = \varkappa (T_s - T_a)\,,
\end{equation}
where $\varkappa$ is the heat transfer coefficient (more correctly, a combination of heat transfer coefficients of different heat transfer modes). A proof of the validity and conditions of acceptance of this approach for biological applications can be found in \cite{Bakken1976}. 

The heat transfer coefficient is a well-known property of heat exchange. Inverse of $\varkappa$ is often used as ``thermal insulance'' in heat engineering applications, as in lumped system analysis, see \cite{Bergman2011}. The correct calculation of $\varkappa$ is a rather complicated task that involves the estimation or modelling of many details of the environment and is outside the scope of this article. An excellent introduction to this matter from the point of view of biology can be found in \cite{Denny1995}.

With such representation of $J_a$ it is clear that temperature variances are connected as
\begin{equation}\label{eq:shell_temperature_varT}
\delta T_s = \tau \dfrac{\varkappa}{\rho_s c_s} \delta T_a\,.
\end{equation}
Due to a steady state definition ($J_a = J_c$) and a constant value of $J_c$, it is clear that, by the end of the transient process, $\delta T_s = \delta T_a$. Of course, this is correct only if equation~\eqref{eq:shell_simple_Ja_linear} is feasible and it is better to say that $\delta T_s$ is a linear function of $\delta T_a$.

This gives us the following estimation of $\tau$:
\begin{equation}\label{eq:shell_tau_est}
\begin{split}
\tau &= \dfrac{\rho_s c_s}{\varkappa} \quad \text{in per unit volume per unit area units, or} \\
\tau &= \dfrac{C_s}{\varkappa A_s} \quad \text{in metric units, or} \\
\tau &\sim \dfrac{\rho_s c_s}{\varkappa} L_s \quad \text{as an order of magnitude estimation}\,,
\end{split}
\end{equation}
where $\rho_s$ is a shell density, $c_s$ is a specific heat capacity of the shell, $\kappa$ is a heat transfer coefficient of the environment, $C_s$ is a heat capacity (bulk) of the shell, $A_s$ is a surface area of the shell, $L_s$ is a ``characteristic size'' of the shell. 

Estimation~\eqref{eq:shell_tau_est} is far from numerical precision and simply demonstrates the dependency of transient process life time from the material properties of the shell and the environment. Nevertheless, the estimation of $\tau$ in \eqref{eq:shell_tau_est} is equal to ``characteristic time'' in solution of Newton's cooling equation, although it has been calculated in completely different way. 

\subsubsection{Formalization completion}
Having results presented by equations \eqref{eq:shell_temperature_varJ} and \eqref{eq:shell_temperature_varT}, one can return to the steady state shell temperature distribution problem \eqref{eq:shell_heat_balance_temperature} and re-define it in more detail:
\begin{equation}\label{eq:shell_temperature_distr}
\begin{split}
&\nabla^2 T_s = 0 \,, \\
&\left. k_s \nabla T_s \cdot \vec{\textbf{n}} \right|_{\partial S} = -J_c\,, \\
&\left. T_s \right|_{\partial A} = f\left( J_a \right) = T_{\textrm{outer}}\left( J_c, T_a \right) = const\,,
\end{split}
\end{equation}
where temperature $T_{\textrm{outer}}\left( J_c, T_a \right)$ is a linear function of $J_a$ and $T_a$ as it follows from equation~\eqref{eq:shell_temperature_varJ}. In the simple case which is defined by equation~\eqref{eq:shell_simple_Ja_linear} and where the Newton's cooling description is appropriate, this function can be expressed as $\left. T_s \right|_{\partial A} = T_a + \frac{1}{\varkappa} J_c $.

The problem \eqref{eq:shell_temperature_distr} can be solved for the same one-dimensional model of the shell which is used for entropy production rate calculation; see the commentaries to equation \eqref{eq:shell_entropyrate_full_1D} for the details of the coordinate system definition. 
\begin{equation}\label{eq:shell_temperature_distr_solution_1}
\begin{split}
&T_s(x) = C_1 x + C_2 \,, \\
&C_1 = -\dfrac{1}{k_s}J_c \,, \\
&C_2 = T_{\textrm{outer}}\left( J_c, T_a \right) - C_1\frac{L}{2} \,,
\end{split}
\end{equation}
where $C_1$ and $C_2$ are integration constants derived from the boundary conditions. A complete one-dimensional solution of steady state temperature distribution in the shell can be presented in the following form:
\begin{equation}\label{eq:shell_temperature_distr_solution}
T_s(x) = T_{\textrm{outer}}\left( J_c, T_a \right) + \frac{1}{k_s}J_c \left( -x + \frac{L}{2} \right) \,,
\end{equation}
and temperature at inner surface $\partial S$ is
\begin{equation}\label{eq:shell_temperature_at_core}
T_s\left(-\dfrac{L}{2}\right) = T_{\textrm{outer}}\left( J_c, T_a \right) + \dfrac{J_c}{k_s} L\,.
\end{equation}

For the simple case where the Newton's cooling approximation is valid, this equation can be written as
\begin{equation}\label{eq:shell_temperature_at_core_N}
T_s\left(-\dfrac{L}{2}\right) = T_a + \left( \dfrac{1}{\varkappa} + \dfrac{L}{k_s} \right)J_c\,.
\end{equation}
Since $L$, $k_s$ and $\varkappa$ are all positive, and $J_c$ is non-negative, one can see that $T_s \geqslant T_a$ where the equality is reached if $J_c = 0$. An illustration of this solution is presented in figure~\ref{fig:shell_temperature_model}.

The temperature of the shell should not exceed the temperature of the core ($T_c \geqslant T_a$), and we can write the condition of ``maximal heat flow throughput'' of the shell:
\begin{equation}\label{eq:shell_flow_throughput}
\begin{split}
&J_c \leqslant \dfrac{k_s}{ L} \left( T_c - T_{\textrm{outer}}\right)\quad \text{or, for Newton's cooling}\\
&J_c \leqslant \left( \dfrac{1}{k_s} L + \dfrac{ 1}{\varkappa} \right)^{-1}\left( T_c - T_a\right)\,,
\end{split}
\end{equation}
and equivalent ``maximal tolerable ambient temperature'':
\begin{equation}\label{eq:shell_max_Ta}
\begin{split}
&T_{\textrm{outer}} \leqslant T_c - \dfrac{L}{k_s}J_c\quad \text{or, for Newton's cooling}\\ 
&T_a \leqslant T_c - \left( \dfrac{1}{k_s} L + \dfrac{ 1}{\varkappa} \right)J_c\,.
\end{split}
\end{equation}

\section{Results Discussion}
\subsection{Shell surface temperature}
The model described does not imply any global dependencies, it works locally and independently in every part of the system (i.\,e. the animal's body). It means that the shell surface temperature may be very inhomogeneous, because of differences in local heat transfer conditions and, possibly, different local core temperature and heat balance requirements.

The model describes the heat transfer only across the shell surface, but heat transfer also works along the shell surface, thus any local temperature excess initiates some lateral heat flows. The model concludes that a shell as a whole should not be described by a stationary equation, as in~\eqref{eq:shell_heat_balance_temperature}, and one should use the full form of the heat transfer equation, including time derivative, spatial dependency of thermal properties, and time dependency in the boundary conditions:
\begin{equation}\label{eq:shell_heat_balance_temp_time}
\begin{split}
&\dfrac{\partial T_s}{\partial t} - 
\nabla \cdot \left( k_s\left( \vec{\textbf{r}}\right) \nabla T_s \right) = 0\,, \\
&\left. k_s\left( \vec{\textbf{r}}\right) \nabla T_s \cdot \vec{\textbf{n}} \right|_{\partial S} = -J_c\left( \vec{\textbf{r}}, t\right) \,, \\
&\left. T_s \right|_{\partial A} = 
T_{\textrm{outer}}\left( J_c\left( \vec{\textbf{r}}, t\right), T_a\left( \vec{\textbf{r}}, t\right) \right) \,,
\end{split}
\end{equation}
where the function $T_{\textrm{outer}}$ is defined by equations~\eqref{eq:shell_temperature_varJ} or~\eqref{eq:shell_simple_Ja_linear}, and is not always known exactly. A solution of equation~\eqref{eq:shell_temperature_distr} can be used as initial conditions.

The process of analytical (or numerical) solution of this equation is incredibly laborious. 

For practical temperature measurement, it means that the shell surface temperature has no obvious connection to the core temperature. For example, attempts to correlate the skin temperature and rectal temperature will mostly fail. 

In addition, equation~\eqref{eq:shell_heat_balance_temp_time} demonstrates that the average surface temperature is weakly defined on a large scale (Laplacian $\Delta T$ can be far from zero).

Nevertheless, it is possible to say that in non-extreme conditions the animal maintains a more or less constant difference between its skin temperature and the temperature of the environment. The value of this difference depends on the activity level of the animal, as well as on some other factors such as health conditions and stress level (i.\,e. depends on the current level of metabolism rate). The higher the activity level, the bigger the difference.

Evidence of linear dependency between ambient and skin surface temperatures was observed in \cite{Loughmiller2001}, where a strong linear correlation was directly measured. Another example of the linear dependence and high inhomogeneity of the skin temperature distribution can be found in \cite{Ward1999}. 

Authors of \cite{Heath2001} found that sheared alpacas had a lower surface temperature than non-sheared. Such an effect can be explained by equation \eqref{eq:shell_temperature_varJ} that predicts surface temperature decrease with an increase of the heat flow into the ambience. The heat flow from the alpacas' skin was increased because shearing significantly decreased the thermo-insulation of the skin (in terms of Newton's cooling, it increases the $\varkappa$ figure). 

\subsection{Newton's law of cooling}
Newton's law of cooling is a common method to describe convective cooling \cite{Bergman2011, Venkanna, Tracy1972}. It simply says that ``the rate of heat loss of a body is proportional to the temperature difference between the body and its environment''. Formal expression of the law is presented by equation \eqref{eq:shell_simple_Ja_linear}. It is important to understand that the law is an approximation, and even a rather empirical relationship. It is not always applicable, and it's feasibility has some implicit assumptions:
\begin{list}{--}{\leftmargin=15mm \labelwidth=5mm \topsep=0mm \labelsep=2mm \itemsep=1pt \parsep=0mm \itemindent=-15pt}
\item temperature difference between the body and environment does not depend on which part of the body is selected for the temperature measurement,
\item the body has a large heat capacity and high thermal conductivity comparing to its environment. Thus, the heat transfer rate inside the body is significantly higher than across the boundary. Sometimes the body can be described as ``an onion'' of region of interest, or ``lumps'', but this should be true inside any region,
\item the heat transfer coefficient depends neither on temperature, nor on time.
\end{list}
If some of these assumptions are not valid, the law may not apply, as is often observed in the presence of free convection (where the coefficient depends on temperature), or during the transition of flow mode (e.\,g. from a laminar to a turbulent one).

Newton's law application in biology is a well-established and rather successful in practice \cite{Bakken1976, SCHOLANDER1950a}. Nevertheless, it should be applied with care, see \cite{Tracy1972}. 

Known values of thermal properties of animal tissues can be found in \cite{Bowman1975} and in \cite[Appendix A]{Kreith1999}. They can be ``averaged'' and the results are presented in the following table (the values are approximate!), where the core is assumed to be composed from blood, muscles and bones, and the shell is presented as built from fat and skin.
\begin{table}[H]
\caption{Thermal properties of tissues}
\label{tab:thermal_prop_media}
\centering
\begin{tabular}{ l c c}
 Media & Specific heat capacity $\left[ \frac{J}{kg \cdot K}\right]$ & Thermal conductivity $\left[ \frac{J}{sec \cdot m \cdot K}\right]$\\
 \hline
 Core & 3700 & 0.50 \\
 Shell & 3000 & 0.20 \\
 Air & 1000 & 0.03 \\
 Water & 4200 & 0.60 \\
 \hline
\end{tabular}
\end{table}

For terrestrial animals the assumption of high body conductivity seems to be valid due to very low heat capacity and the thermal conductivity of the air, but for aquatic animals it's validity can be arguable. 

The model proposed in this article does not require such assumptions regarding thermal properties of the animal's body. It uses different assumptions (see above), where one of the most important constraints is requirement of slow changes in the environment to enable the shell to stay in a steady state.

If this assumption is valid, the model embodies Newton's cooling as a special case (see equations~(\ref{eq:shell_simple_Ja_linear} -- \ref{eq:shell_tau_est})). In the model's framework, the Newton's cooling law can be interpreted as an empirical description of transient processes between different steady states of the shell. For example, if the temperature of the environment is lowered then the skin surface temperature follows it according Newton's law of cooling with apparent characteristic time given by equation \eqref{eq:shell_tau_est}. The same is true if a transition process has been initiated by the core (change in heat flow because of metabolism rate variation). For example, if the animal ceases an activity (i.\,e. lowers the metabolism rate), the skin surface cools according to Newton's law.

When the shell is in a steady state, the Newton's law of cooling may not always be an adequate mechanism if one wants to correlate core temperature, skin temperature, and temperature of the ambience using measured or estimated thermal properties of the body and the media. 

\subsection{Regulation limitations}
The major claim of the model is the hypothesis concerning the role of blood flow in thermoregulation of animals. It is absolutely clear that thermoregulation in not a primary purpose of the blood, and that the temperature regulation is not a top priority task of the blood flow control system. 

The blood flow velocity lies inside a specific range, and this range fits the primary purposes of the blood system. From the point of view of thermoregulation, this is a limitation: 
\begin{equation}\label{eq:regulation_blood_minmax}
v_{\min} \leqslant \left|\vec{\textbf{v}}\right| \leqslant v_{\max}\,.
\end{equation}
Indeed, if $v=v_{\min}$ already, but $\left.\nabla T_c\right|_{\partial S}$ continues to increase (permanent cooling of the ambience) then it can not continue to be compensated for. A similar problem rises at the high limit of the flow velocity range $v=v_{\max}$, as can be explained by the following regulation equation:
\begin{equation}\label{eq:regulation_control}
\left.\left(-k_c \nabla T_c + \rho \vec{\textbf{v}} c_v T_c \right) \cdot \vec{\textbf{n}} \, \right|_{\partial S}= \left.\vec{\textbf{J}}_c \cdot \vec{\textbf{n}} \, \right|_{\partial S}\,.
\end{equation}

Control mechanics may not be perfect, and if the flow velocity is set too low (at low ambient temperature) then the shell may suffer from frost blight. A high temperature of the ambience and corresponding high flow velocity may cause abnormal pressure in vessels exceeding their breaking point, thus haemorrhages may occur in the shell. There are researches which demonstrate that vasculature could be built in a sophisticated way to relax these limitations. See \cite{Ninomiya2011, SCHOLANDER1950a, Coffman1966}.

Another limitation of the regulation is connected to the maximum shell throughput (see~\eqref{eq:shell_flow_throughput} and~\eqref{eq:shell_max_Ta}):
\begin{equation}\label{eq:regulation_flow_max}
J_c \leqslant J_{\max} = \dfrac{k_s}{ L} \left( T_c - T_{\textrm{outer}}\right)\,.
\end{equation}
This relation demonstrates that the heat flow from the core can not be set arbitrarily high, and that the upper limit depends on the difference between core temperature and temperature of the environment: the lower the difference then the lower the maximum acceptable value of the heat flow $J_{\max}$, see equation~\eqref{eq:shell_temperature_distr} for more details about $T_{\textrm{outer}}$. Some signs of this limitation can be found in \cite{Ward1999} in the data related to the areas with highest measured heat flow. 

This limitation is especially important if an ambient temperature is high. At high temperature and high metabolism rate (when high $J_c$ is required), the shell may saturate the heat flow, and the core will be forced to use additional mechanisms of the thermoregulation if possible, and/or lower the metabolism rate. For example, cease chasing the prey and start panting. This ``heat sink bottleneck'' problem may give additional explanation to some effects that can be observed during work or exercise in hot environments, see \cite{Casa1999, Walters2000, Speakman2010, Brake2002, Epstein2006}. 

Equation~\eqref{eq:shell_max_Ta} means that for every $J_c$, or, equivalently, for every metabolism rate, there is a maximum of tolerable (operational) temperature of the environment. Note that it is better to speak about the heat flows, instead of the temperatures, because this limitation depends on thermal properties of both the body and the media, and upper limit of the temperature in air may differ from the limit in water \cite{Denny1995}.

\subsection{Adaptation to climate}
Most, if not all, of the regulation mechanics and corresponding limitations depend on thermal properties of body tissues, and some depend on body dimensions. The thermal properties ($\rho$, $C$, $k$) form ``hard'' limitations. One can hardly expect significant change in heat capacity, thermal conductivity, or density of the body in a reasonable time that is necessary for the proper thermoregulation. 

Other properties are more flexible: for instance, $\varkappa$, which defines ``insulation'' of the body surface can be cleverly engineered using fur, feathers, behavioural patterns (i.\,e. staying in shade to reduce radiation heating, walking to the wind during a hot day to increase forced convection, etc). See, for example, \cite{Fratto2011, Stockman2006, Myhrvold2012}.

Comprehensive investigation of animal adaptation to cold and hot environments in regard to metabolism rate, body temperature, and temperature of the ambience can be found in \cite{SCHOLANDER1950, SCHOLANDER1950c}. From the proposed model's point of view, the adaptation can be done in some additional ways explained below.

As the regulation equation \eqref{eq:regulation_control} says, the ``initial'' value of the temperature gradient $\left.\nabla T_c\right|_{\partial S}$ can be chosen as appropriate for the expected temperature range of the environment. Thus, in cold climate, where low $T_a$ is quite probable, lower value of the gradient allows the blood flow velocity to be set at a higher level, preventing frost blight. It can be achieved by increasing subcutaneous thickness (e.\,g. using thicker layer of fat) to limit the steepness of the temperature change, and/or by increasing skin thickness, which, according to the equation~\eqref{eq:shell_temperature_at_core}, results in raising of the inner temperature of the skin. In a cold climate, thick skin and subcutaneous tissues allow high level of metabolism rate and high blood circulation rate without unnecessary heat loss. Evidence of the vascular control's role in body temperature regulation in cold climate can be found in \cite{SCHOLANDER1950a, Olnianskaya1947}.

In a hot climate with a high probability of high ambience temperature, this approach should be reversed: higher $\left.\nabla T_c\right|_{\partial S}$ leaves more freedom in blood flow regulation. Thinner skin avoids the ``heat flow saturation'', as can be seen from the equation~\eqref{eq:regulation_flow_max}. In a hot climate the thin skin and subcutaneous tissues are beneficial if high level of metabolism rate is required. 

These adaptation and regulation mechanisms do not include those related to the core temperature choice, metabolism rate regulation, or mechanisms that are important when the ambient temperature is far outside of the thermo neutral zone of the animal, which are not considered in this article and are beyond the scope of the proposed model.

\section{The model and beyond}
\subsection{The model and reality}
Homoeothermal animals differ from any other creatures (either animals, or plants, or fungi) with their ability to maintain the inner temperature of the organism. This ability is absolutely crucial from the point of view of an implementation of a feasible model of the thermoregulation of a being. Endothermic animals have a core that permanently generates heat energy because of metabolism. Thus, a model of thermoregulation must include a heat production and a heat discharge with ways for the latter to be optimal; this forces a researcher to take into consideration an entropy production and entropy discharge, beside the heat production modelling.

The model described above implies a homogeneity of a core; that is supposed to remain in a (quasi)equilib-rium: there are no inner heat flows, or entropy flows. From a biological point of view, such species are the \textsl{homoeotherms}\footnote{And man is among them: this fact stands behind the body temperature measurement as a diagnostically valuable medical procedure.}. Alternatively, some species are known for a significant variation of the temperature of the compartments of the body; such animals are called \textsl{heterothermic}. The core temperature ($T_c$) of homoeotherms should be constant for optimal performance of enzymes regulating core biochemistry. It holds true for heterotherms also with slight modification to \textit{mostly constant}.

According the the First Law of Thermodynamics this requirement leads to the need of a heat flow ($J_c$) that dissipates excessive power to the external environment. The power that should be dissipated is proportional to the current metabolism rate value. Constant $T_c$ requirement implies constant $J_c$ with a power that is appropriate for the given metabolism rate.

If the core were directly exposed to the environment it would be strongly driven by ambient temperature (as well as other thermal parameters) fluctuations but would not possess enough regulation options. That is why the shell always exists, both in the model and in reality. Thermodynamic parameters regulation means variable material properties (heat capacity, thermal conductivity, density, etc). It leads to obvious difficulties if an immediate response to significant ambience variation is required.

Under the First Law, some work may be involved to control heat gain or loss ($\mathrm{d}U = \mathrm{d}Q + p \, \mathrm{d}V$), but this may fail to meet other constraints that are not related to the thermal balance. For example, variable volume may involve use of some thin elastic constructions that may not provide serious mechanical protection.

Simultaneously, a notable mass exchange ($\mathrm{d}U = \mathrm{d}Q + p \, \mathrm{d}V + \mu \, \mathrm{d}N$) may not provide a solution of the problem: both mass and composition of the core of any creature are usually permanent. This conservation of course is not absolute: food/water consumption, growth, etc. make some slight variations to those figures. Yet, the variations are too small to be effective for quick responding to a temperature change.

These two issues result in a shell implementation, with no ambiguity. The shell acts as a heat buffer that reduces variations of thermal parameters of the environment as they are experienced by the core. When ambience warms up, the shell absorbs the excess head coming from the body (shell entropy increases). Reciprocally, when the ambience cools down the shell radiates some heat (shell entropy decreases) removing the burden of regulation from the core.

The shell is integral part of the body, so the implementation of the mechanisms for heat exchange between the shell and the core seems to be rather simple. Because the shell temperature ($T_s$) is set up ``automagically'' by the Second Law of Thermodynamics and is not controlled by the core, it is much easier to achieve other ``design goals'' (e.\,g. mechanical protection), because the constraints to the thermal parameters of the skin may be relaxed (compared to the constraints in the case of the directly connected core). All mechanics described in this article can work locally. It means, that it is possible to set $T_c$ and $J_c$ as it is optimal locally, and the shell will find a steady state as is appropriate for the given $T_a$ (more strictly, for a given local $J_a$). This means that shell temperature is not required to be homogeneous over the entire surface of the animal.

The model presented here is basically aimed at describing the processes of a temperature regulation for homoeothermic animals: it implies a linear approximation of some processes, a significant difference between an organism and an ambience. Thus, it hardly could be extended for the poikilothermic species, or plants, immediately. Nonetheless, there could be developed versions of the model for these types of organisms, while the discussion of this issue falls beyond the scope of this paper.

\subsection{Beyond the model}
Yet, the model presented in the article may surely be improved and extended; here we discuss some possible extensions for consideration.

\subsubsection{Poikilothermic species}
There are no plant or fungi species that possess maintenance of permanent body temperature; nor invertebrates. Fishes, Amphibians and Reptiles are all poikilothermic ones. Surprisingly, there is a single species among Mammalians (the naked mole rat \textsl{Heterocephalus glaber}) that is a poikilothermic species. It has a number of specific peculiarities that seem to be quite incompatible to the poikilothermism: very complex social structure of family, very strong stability against a thermal or a chemical burn, etc. All these features are to be hypothesized to get a relation to its poikilothermism.

Poikilothermic species are much more endangered by overheating, in comparison to the homoeothermal animals. Since they are seriously restricted in the redistribution of their inner energy fluxes in a manner to switch on the (active) mechanisms of an excessive heat discharge. Thus, the model described above must adjust to take into consideration very fine differences in the temperatures of the core and the ambience; this point seems to be rather technical, but a lot of biological issues are expected to stand behind. An extension of the model for plant and/or fungi still requires many specific issues to be taken into consideration.

\subsection{Diurnal rhythms}\label{diur_r}
All mammalians exhibit a two-phase day activity pattern: a sleep and an activity, itself. It is a well-known fact that the metabolism levels in the active state, and in the sleep state differ significantly. One might say that the sleep metabolism is pretty close to a (comfort) minimal one: a number of (regular) physiological functions (e.\,g., urine production in kidneys, muscle work, etc.) are ``around zero'' level. Certainly, a heart, and breathing muscles still continue to work, thus sustaining the comfortable minimum level mentioned above.

Thus, one needs to develop two basically different models of thermal regulation: the former valid during an active period of the day, and the latter valid for the sleep. This brings a researcher to an idea of the implementation of the combined model that is able to describe (and predict) the switching of the essential thermal regulatory mechanisms during transition between these two states of the organism.

\subsection{Hibernation and torpor}
A number of species exhibit another type of two-phase activity pattern, called hibernation and/or torpor. They differ, actually, from the biological point of view; meanwhile, we shall discuss them as an entity, with no differentiation. A torpor differs from the diurnal activity pattern in the figures of the typical temperature, and the metabolic activity exhibited by organisms. In fact, a hibernation is accompanied by a significant (up to ten grades) body temperature decrease, with reciprocal shut down of a metabolism.

Usually, hibernation and torpor are strongly correlated with the seasonal activity of a species; a number of rodents act in this way. Such a life pattern is rather wide spread, and is not restricted only to rodents.

From the technical point of view, a model that combines both the model described above, and another model that is valid in the torpor state, may look similar to the model described in Section~\ref{diur_r}. Essentially, it would differ rather strongly, because the day rhythm metabolism variations and the torpor metabolism variations have drastically different biological natures.

% Bibliography:

\end{document}